# Ammonia and CO observations toward low-luminosity 6.7-GHz methanol masers

Y. W. Wu[1], Y. Xu[1], J. D. Pandian[2,3], J. Yang[1], C. Henkel[3], K. M. Menten[3], S. B. Zhang[1]

## ABSTRACT

To investigate whether distinctions exist between low and high-luminosity Class II 6.7-GHz methanol masers, we have undertaken multi-line mapping observations of various molecular lines, including the $NH_3$ (1,1), (2,2), (3,3), (4,4) and $^{12}CO$ (1-0) transitions, towards a sample of 9 low-luminosity 6.7-GHz masers, and $^{12}CO$ (1-0) observations towards a sample of 8 high-luminosity 6.7-GHz masers, for which we already had $NH_3$ spectral line data. Emission in the $NH_3$ (1,1), (2,2) and (3,3) transitions was detected in 8 out of 9 low-luminosity maser sources, in which 14 cores were identified. We derive densities, column densities, temperatures, core sizes and masses of both low and high-luminosity maser regions. Comparative analysis of the physical quantities reveals marked distinctions between the low-luminosity and high-luminosity groups: in general, cores associated with high-luminosity 6.7-GHz masers are larger and more massive than those traced by low-luminosity 6.7-GHz masers; regions traced by the high-luminosity masers have larger column densities but lower densities than those of the low-luminosity maser regions. Further, strong correlations between 6.7-GHz maser luminosity and $NH_3$ (1,1) and (2,2) line widths are found, indicating that internal motions in high-luminosity maser regions are more energetic than those in low-luminosity maser regions. A $^{12}CO$ (1-0) outflow analysis also shows distinctions in that outflows associated with high-luminosity masers have wider line wings and larger sizes than those associated with low-luminosity masers.

*Subject headings:* masers — ISM: molecules — ISM: jets and outflows — stars: formation

[1]Purple Mountain Observatory, Chinese Academy of Sciences, Nanjing 210008, China

[2]Institute for Astronomy, University of Hawaii, 2680 Woodlawn Drive, Honolulu, HI 96822, USA

[3]Max-Planck-Institut für Radioastronomie, Auf dem Hügel 69, 53121 Bonn, Germany

## 1. Introduction

The 6.7-GHz methanol maser is the second brightest Galactic maser after the 22-GHz $H_2O$ maser (Menten 1991). More importantly, it is only found in massive star-forming regions (e.g. Minier et al. 2003), which makes it a signpost of massive star formation (Ellingsen 2007a). At present, extensive surveys have yielded more than 800 6.7-GHz maser sites (Caswell et al. 1995; Pestalozzi et al. 2005; Pandian et al. 2007a; Ellingsen 2007b; Xu et al. 2008; Caswell 2009; Green et al. 2009). Theoretical models suggested that the masers are produced by radiative pumping in a warm, dense and dusty environment (Sobolev et al. 1997; Cragg et al. 2001, 2005). High resolution observations indicate that only a small percentage are associated with H II regions (Walsh et al. 1998), but all are associated with massive millimeter (Beuther et al. 2002a) and sub-millimeter sources (Walsh et al. 2003). Interferometric observations of ammonia ($\sim 10''$ resolution) towards southern 6.7-GHz masers revealed a strong spatial correlation between methanol masers and ammonia cores (Longmore et al. 2007).

However, some major puzzles remain. Current maser theories focus on high-luminosity masers, such as the one in W3(OH) (Sutton et al. 2001), while low-luminosity masers are more numerous. Szymczak & Kus (2000) found that many low-luminosity methanol masers were associated with IRAS sources whose colors are different from those expected for ultracompact H II regions. While Pandian et al. (2007b) did not find differences based on their multi-wavelength counterpart analysis, recently, Breen et al. (2010) found that low-luminosity 6.7-GHz maser regions are more dense than their high-luminosity counterparts. Thus, the question remains as to whether low-luminosity 6.7-GHz methanol masers have properties different from those of their high-luminosity counterparts.

Hence, comparison of the physical properties of low and high-luminosity methanol masers will be valuable for furthering our understanding of 6.7-GHz methanol masers and their relationship with the early stages of massive stellar evolution. For this purpose, we selected a sample of 9 low-luminosity 6.7-GHz methanol masers, and investigated their physical conditions and kinematic properties using molecular tracers, including the (J,K) = (1,1), (2,2), (3,3), and (4,4) inversion transitions of ammonia and the $J = 1-0$ rotational transition of $^{12}CO$. For comparison, a sample of 8 high-luminosity 6.7-GHz masers with comparable $NH_3$ data was observed in $^{12}CO$ (1-0). As discussed by Ho & Townes (1983), ammonia is an excellent probe of the physical conditions of dense molecular gas. It is a good molecular cloud thermometer (Walmsley & Ungerechts 1983), has a moderately high critical density of $1.9 \times 10^4$ cm$^{-3}$ (Stahler & Palla 2005), and is more resistant to the effects of depletion than other high density tracers such as CS (Bergin & Langer 1997). In contrast, the $^{12}CO$ (1-0) line can be used to probe the kinematics, especially the outflows, of the maser regions.

This paper is organized as follows. In Sect. 2 we describe the samples and observations. In Sect. 3 we present the results including molecular line maps and physical quantities of the regions. Comparative analysis is given in Sect. 4. Finally, conclusions are drawn in Sect. 5.

## 2. Samples and Observations

### 2.1. Samples

The low-luminosity 6.7-GHz methanol masers in this study were mostly selected (7 out of 9) from the catalog of Xu et al. (2003) according to their lowest luminosities (the luminosities are calculated from the peak flux density assuming a typical linewidth of 0.25 km s$^{-1}$ and isotropic emission). Two other sources with very low luminosities, 106.80+5.31 and 206.54-16.36, were selected from Xu et al. (2008) and Minier et al. (2003) respectively. The 6.7-GHz line luminosities of these sources range from $4.4 \times 10^{-10}$ to $1.5 \times 10^{-7}$ L$_{\odot}$. Their properties, including galactic and equatorial coordinates, peak flux densities, distances, luminosities, other names and references are listed in Table 1. The distances for 111.25-0.77, 183.35-0.59 and 189.78+0.34 are kinematic (i.e. derived from their radial velocities), while for the other sources, distances are derived from the absolute magnitudes of O and B stars.

To compare the physical conditions of low and high-luminosity maser regions, we also need a sample of high-luminosity masers. To select a high-luminosity maser sample, we used the following criteria: first, the maser luminosity must be at least an order of magnitude larger than that of the low-luminosity sources. Second, the sources must have previous NH$_3$ (1,1) and (2,2) data that is comparable to our data for the low-luminosity sources. Using these selection rules, we compiled a list of 8 high-luminosity 6.7-GHz masers, tabulated in Table 2. NH$_3$ (1,1) and (2,2) spectral parameters of these sources are also compiled and listed in Table 3. For 12.68-0.18 and 173.49+2.42, their distances are kinematic distances, and for the other sources, their distances are derived from trigonometric parallax.

The low-luminosity sources are all within a distance of 4 kpc from the Sun. The uncertainties in distances are estimated to be less than 30%, even for the kinematic distances since the sources are relatively nearby (Roman-Duval et al. 2009). While the high-luminosity sources are typically more distant than the low-luminosity sources, their distances are mostly determined from trigonometric parallax and are consequently more accurate than that of low-luminosity sources. Hence, we expect the influence of distance errors on our study to be limited.

## 2.2. Observations

The mapping observations of $NH_3$ (1,1), (2,2), (3,3) and (4,4) transitions towards the low-luminosity maser sample were made with the Effelsberg 100-m telescope[1] during December 2007 and March 2008. The system temperatures were $40-90$ K during the observations. The beam size was $40''$ at 23 GHz. The frontend was a dual channel K-band HEMT receiver, while the backend was the AK90 autocorrelator which allowed the $NH_3$ (1,1), (2,2), (3,3) and (4,4) lines to be measured simultaneously. The eight available sections of the AK90 autocorrelator were divided into two groups to sample the two orthogonal, linearly polarized components (Henkel et al. 2000) of each line. The observations were made in frequency switching mode with a frequency throw of 7.5 MHz. The grid spacing of the mapping observations was $30''$, with an integration time of 10 min per grid point. A $2' \times 2'$ region was mapped around each maser position, with the area being expanded suitably to enclose emission with intensity more than 50% of peak emission. The resulting typical map size was $2.5' \times 2.5'$, while the maximum map size was $5' \times 4'$. The pointing was checked every hour on nearby continuum sources and was found to be better than $10''$. W3(OH) was used for flux calibration. We estimate an absolute calibration accuracy of $\pm10\%$.

The $^{12}CO$ (1-0) observations (towards both low and high-luminosity masers) were performed in November 2008 with the 13.7-m millimeter-wave telescope in Delingha, China, which is operated by the Purple Mountain Observatory. A cooled SIS receiver (Zuo et al. 2004) was used for the frontend with system temperatures ranging from 200 to 300 K during the observations. An Acousto-Optical Spectrometer (AOS) was used as the backend. The HPBW was $60''$ at 115 GHz. The observations were performed in a position switched mode. The grid spacing of the mapping observations was $30''$, the average integration time per point 5 min. The pointing accuracy was better than $10''$. Data were calibrated using the standard chopper wheel method. Absolute calibration is estimated to be accurate to about 15%. Table 4 shows a summary of our observations.

The $NH_3$ and CO data were processed using the CLASS and GREG packages of the GILDAS[2] software.

---

[1] Based on observations with the 100-m telescope of MPIfR (Max-Planck-Institut für Radioastronomie) at Effelsberg.

[2] CLASS and GREG are part of the Grenoble Image and Line Data Analysis Software (GILDAS) working group software. http://www.iram.fr/IRAMFR/GILDAS/

## 3.  Results and Analysis

NH$_3$ (1,1), (2,2) and (3,3) emission was detected in all low-luminosity sources at the 3$\sigma$ level of 0.45 K except for 133.72+1.22. While no NH$_3$ (4,4) emission was found at the 3$\sigma$ level of 0.36 K. The NH$_3$ (1,1) spectra were fitted using "METHOD NH3(1,1)" in CLASS, which fits the hyperfine structure and derives optical depths and line widths accounting for line broadening from optical depth effects. The NH$_3$(2,2) and (3,3) lines were fitted with Gaussian profiles. In Fig. 1 we presents the NH$_3$ (1,1), (2,2) and (3,3) spectra at the positions of the peak of the ammonia cores with the fit profiles overlaid on the data. The fit parameters, i.e., the brightness temperature, line width, the radial velocity with respect to the local standard of rest (V$_{LSR}$), optical depth of the NH$_3$ (1,1) line and integrated intensities are listed in Table 5.

Fig. 2 shows contour maps of NH$_3$ (1,1), (2,2) and (3,3) emission overlaid on the Midcourse Space Experiment (MSX) E band (21 $\mu$m) image. For 106.80+5.31, where no MSX data are available, the Multiband Imaging Photometer for Spitzer (MIPS) 24 $\mu$m image is used. The open squares, filled triangles and ellipses in Fig. 2 denote the methanol masers, the MSX point sources and error ellipses of the IRAS point sources in the fields, respectively. The NH$_3$ (1,1) maps were used to identify cores, and determine their sizes and positions with the exception of source 106.80+5.31 where the cores were characterized by their NH$_3$(2,2) emission. Cores were identified using visual inspection. The methodology used to define cores is similar to the operation of the "CLUMPFIND" algorithm (Williams et al. 1994). In total, We identified 14 cores from the 8 fields, with 6 of them lacking maser associations. Although our observations were spatially undersampled, the cores detected in our observations have sizes well above the beam size, and we do not expect the undersampling to have a big impact on our study.

We use the $^{12}$CO (1-0) data mainly for identifying and analyzing outflows. We find outflows to be common among both low and high-luminosity methanol masers. Among the low-luminosity maser regions, we find 6 outflows in 9 regions. For the high-luminosity maser regions, we find outflows in four sources. For the other sources, 9.62+0.20, 12.68-0.18, 35.20-1.74 and 49.49-0.39, the spectra have multiple velocity components, making it impossible to identify line wings, and hence outflows. Fig. 3 shows the $^{12}$CO (1-0) spectra for the ten sources with detected outflows, and Fig. 4 shows maps of the outflows.

### 3.1. Physical Quantities

We obtained physical parameters of sources from the inversion transitions of $NH_3$. Excitation temperatures of $NH_3$ (1,1), $T_{ex}$ (1,1), and ammonia column densities, $N(NH_3)$, were obtained using the formulae in Harju et al. (1993). Rotational temperatures between the (1,1) and (2,2) levels, $T_{R21}$, and molecular hydrogen densities, $n(H_2)$, were calculated using the equations given by Ho & Townes (1983). Kinetic temperatures were derived following Walmsley & Ungerechts (1983), modified with the new collision rate coefficients of Danby et al. (1988). The column density of molecular hydrogen, $N(H_2)$, can be obtained from the $NH_3$ column density by assuming the fractional abundance of $NH_3$. Here, we adopted a value of $3 \times 10^{-8}$ for $[NH_3]/[H_2]$ (Harju et al. 1993). It should be noted that there is a significant uncertainty in the $NH_3$ abundance (which is potentially variable within a cloud), which translates to an uncertainty in the $H_2$ column density. We discuss this in more detail in §4.3.

The nominal core sizes, $l$, were determined by de-convolving the telescope beam, using equation (1),

$$l = D \left( \theta_{1/2}^2 - \theta_{MB}^2 \right)^{1/2},$$

where $D$ is the distance, $\theta_{1/2}$ is the half-power angular size of the core, and $\theta_{MB}$ is the half-power beam width of the telescope. Core masses were computed by assuming a Gaussian column density distribution with a full-width at half-maximum of $l$. If $N(H_2)$ is the molecular hydrogen column density of peak position, $m_{H_2}$ is the mass of the hydrogen molecule and $\mu$ is the ratio of total gas mass to hydrogen mass (assumed to be 1.36 based on Hildebrand 1983), the core mass is given by

$$M \simeq \mu m_{H_2} \int 2\pi r N\left(H_2\right) e^{-4\ln 2\ (r/l)^2} dr,$$

The derived core sizes, the kinetic temperatures, excitation temperatures, molecular hydrogen densities, column densities and masses of the 14 cores identified in our observations are listed in Tables 6. Though the physical quantities of regions associated with high-luminosity masers are available in the literature, we recomputed them using the same methodology we used for the low-luminosity masers for consistency. Since the optical depth of the source 188.95+0.89 was not given in the literature, we solved it from equation (4) of Ho & Townes (1983) :

$$T_{R21} = -41.5 \div \ln \left[ \frac{-0.282}{\tau_m(1,1)} \ln \left\{ 1 - \frac{\Delta T_a^*(2,2,m)}{\Delta T_a^*(1,1,m)} \times (1 - e^{-\tau_m(1,1)}) \right\} \right].$$

The physical parameters of cores associated with high-luminosity masers are listed in Table 7.

The properties of the outflows were estimated using the method of Beuther et al. (2002b), except for the $^{12}$CO column density which was derived from Snell et al. (1988). The CO excitation temperature and the $[^{12}CO]/[H_2]$ abundance ratio adopted were 30 K and $10^{-4}$, respectively. Typically, the molecular outflows have masses of a few solar mass, momenta of tens of $M_\odot$ km s$^{-1}$, kinetic energies of $\sim 10^{45}$ ergs, mass entrainment rates of a few $10^{-5}$ $M_\odot$ yr$^{-1}$, mechanical forces of a few $10^{-4}$ $M_\odot$ km s$^{-1}$ yr$^{-1}$, and mechanical luminosities of about 1 $L_\odot$, respectively. The characteristic time scale is equaling the radius divided by the maximum velocity separation of the red and the blue wings, are $10^4 - 10^5$ years.

As we assumed the line wings to be optically thin, quantities involving masses, mass loss rate, mechanical force and mechanical luminosity are lower limits to the true values. The observational parameters of the outflow wings, including LSR velocity ranges, integrated fluxes and full width at 100 mK levels are listed in Table 8. The physical properties of the outflows, including masses, sizes, the characteristic time scales, $t$, the mass loss rates, $\dot{M}_{out}$, the mechanical forces, $F_m$, and the mechanical luminosities, $L_m$, are summarized in Table 9.

### 3.2. Comments on Individual Sources

*106.80+5.31* — This source is associated with the central part of the S140 molecular complex. NH$_3$ emission shows filamentary structure extending southwest to northeast. The peak position of NH$_3$ (1,1) emission is offset $\sim 40''$ northeast of the mid-infrared peak while the peak position of NH$_3$(3,3) emission is located at the southwest edge of the mid-infrared dust emission. The NH$_3$(2,2) map reveals a two-core structure, with the northeastern core consistent with the NH$_3$ (1,1) peak and the southwestern core consistent with the NH$_3$(3,3) peak. High resolution NH$_3$ (1,1) and (2,2) maps (see Zhou et al. 1993, Fig. 1) show a molecular cavity surrounded by warm dense gas traced by NH$_3$.

*111.25-0.77* — There is a core that peaks towards IRAS 23139+5939. NH$_3$ observations give a kinetic temperature of 31 K and mass of 40 $M_\odot$.

*121.24-0.34* — There is a core elongated from the northwest to southeast and peaks at the location of the IRAS source 00338+6312. Temperature and line width gradients are found, with the kinetic temperature and NH$_3$ (1,1) line width decreasing northwest to southeast from 21 K, 2 km s$^{-1}$ to 13 K, 1 km s$^{-1}$.

*133.72+1.22* — This is an intensively studied region of active star formation, the W3 complex. There is significant evidence revealing a newly formed star cluster in this complex, e.g., several radio continuum components (Colley 1980) and compact and diffuse infrared sources (Wynn-Williams et al. 1972). However, We did not detect NH$_3$ emission in this region, which may due to our limited sensitivity.

*183.35-0.59* — A single NH$_3$ core is located southwest and with offset of $\sim 15''$ from the IRAS source 05480+2545. Klein et al. (2005) found a dust core through millimeter continuum with an identical morphology to the NH$_3$ core.

*188.80+1.03* — A filament consisting of two cores extends southwest to northeast through the field. Core 1, which is offset by $\sim 70''$ north-east to the IRAS source 06061+2151 is more massive and warmer than Core 2.

*189.03+0.78* — A filament consisting of two cores extends southwest to northeast through the field. Core 1, the southwest core, is associated with IRAS 06056+2131. It has an angular extent of 80$''$ and peaks at a mid-infrared MSX point source. Core 2, the northeast core, has an angular extent of 70$''$ with its peak offset $\sim 15''$ from another mid-infrared MSX source.

*189.78+0.34* — This source is part of the S252A molecular complex. NH$_3$ maps reveal a filamentary structure, composed of three cores, extending from northwest to southeast. Core 1, associated with a methanol maser, is situated $\sim 30''$ southeast to IRAS 06055+2039. Cores 2 and 3, with separations of $\sim 90''$ and 180$''$ from Core 1 respectively, reside in the middle and the southern parts of the filament.

*206.54-16.36* — It is located in the vicinity of NGC 2024 HII region. A filament extending in the north-south direction, called "Molecular Ridge" exists in this region (Chandler & Carlstrom 1996). The filament resolves into 2 segments in the NH$_3$ (1,1) and NH$_3$(2,2) maps. Along the ridge there are seven far infrared sources (Mezger et al. 1988, 1992). The CH$_3$OH maser is associated with FIR4 (Minier et al. 2003).

## 4. Discussion

Here, we present a comparative analysis of the physical quantities, including spectral line width, temperature, density, column density, core size, mass and outflow properties between the low-luminosity and high-luminosity methanol masers.

### 4.1. Temperature

Fig. 5 show the NH$_3$(1,1) excitation temperature, T$_{ex}$ and the kinetic temperature, T$_{kin}$, as a function of maser luminosity. It is seen that the excitation temperature of NH$_3$(1,1) line of low-luminosity maser regions is on average larger than that of high-luminosity maser regions. The mean value of T$_{ex}$(1,1) for the low-luminosity maser regions is 10 K, which is 4 K larger than that of the low-luminosity maser regions. On the other hand, there is no distinction in the kinetic temperature between low- and high-luminosity masers. To test the statistical significance of these results, we conducted a two sample t-test with the hypothesis of same mean values for the two groups to determine if the mean values of the two groups is different. The t-test yields a p value of 0.0016 for $T_{ex}$ and 0.12 for $T_{kin}$. The p-value is the probability that one would measure a discrepancy that is greater than what is observed, assuming the null hypothesis to be true. Generally, if the value of p is smaller than 0.05 or 0.01, we can conclude at the 95% of 99% confidence level that the null hypothesis can be rejected. Thus we conclude that the difference in $T_{ex}$ between the two groups is statistically significant while there seems to be no significant difference in $T_{kin}$ between the two groups. As $T_{ex}(1,1)$ is determined by $n(H_2)$, $T_{kin}$ and local radiation field, the lower $T_{ex}(1,1)$ of the high-luminosity maser regions may rise from lower densities in these regions (see Ho et al. 1977, Fig. 4.4). Subsequent density comparisons confirm this opinion.

## 4.2. Line width

For the low-luminosity methanol masers, the $NH_3$ (1,1) line widths range from 1.22 to 2.72 km s$^{-1}$ with a mean value of 1.75 km s$^{-1}$, while the $NH_3$ (2,2) line widths range from 1.24 to 2.83 km s$^{-1}$ with a mean value of 1.77 km s$^{-1}$. In contrast, the $NH_3$ (1,1) line widths range from 2.16 to 8.8 km s$^{-1}$ with a mean value of 3.9 km s$^{-1}$, and $NH_3$ (2,2) line widths range from 1.9 to 8.3 km s$^{-1}$ with a mean value of 4.1 km s$^{-1}$ for the high-luminosity methanol masers. It is clear that the high-luminosity maser group exhibits larger line widths compared to low-luminosity masers.

To explore this further, we plot the $NH_3$ (1,1) and (2,2) line widths as a function of the maser luminosity (Fig. 6). Fig. 6 shows a correlation between the maser luminosity and the line widths of both (1,1) and (2,2) lines. A least-squares fit yields power-law slope of 0.14 for both $NH_3$ (1,1) and (2,2) line widths. The intercepts for both fit are 1.28 and 1.27, which is roughly the same and corresponding correlation coefficients, R, are 0.81 and 0.83 respectively. The similarity between these two fits is interesting. This, when considered with the similar line widths of the (1,1) and (2,2) lines, suggests that both lines are tracing identical regions of the cores, contrary to what was found in infrared dark clouds by Pillai et al. (2006). This is confirmed when inspecting the integrated intensity maps for $NH_3$ (1,1) and (2,2) emission, since morphology for these two emission is very similar.

Line widths are considered to be indicators of internal motions in the interstellar medium (Larson 1981). The observed linewidths may arise from bulk motions such as rotation or turbulence. Although it is not possible to distinguish these components with our observa-tions, the correlation found here indicates that, generally, internal motions in high-luminosity maser regions are more energetic than their low-luminosity counterparts.

### 4.3. Density and Column Density

Differences in densities and column densities between the two groups are also found. The mean and median values of $n(H_2)$ for the low-luminosity methanol masers are 1.9 and 1.8 $\times 10^4$ cm$^{-3}$, three times larger than that of the high-luminosity masers (see Table 6, 7). This is corroborated by similar results based on 1.2 mm dust continuum observations towards methanol masers in the southern sky (Breen et al. 2010).

In contrast, the column densities show an opposite relationship: the high-luminosity maser regions have larger column densities. Mean and median values of $N(NH_3)$ of the high-luminosity group are 1.9 and 1.5 $\times 10^{15}$ cm$^{-2}$, which are three times larger than those of the low-luminosity group. Fig. 7, which shows the $H_2$ density and $NH_3$ column density as a function of maser luminosity, shows the dichotomy between the two groups. It is particularly interesting to see a gentle increase of column density with the maser luminosity less than $10^{-6}$ L$_\odot$ followed a sharp jump of column density when the maser luminosity becomes larger than than $10^{-6}$ L$_\odot$. A two sample t-test with the hypothesis of same mean values for the two groups yields p values of 0.0016 and 0.017 for $n(H_2)$ and $N(NH_3)$, indicating that the density and column density differences between the two groups are all significant.

Habitually, the molecular hydrogen column density, $N(H_2)$, can be obtained from the ammonia column density, $N(NH_3)$, by assuming the ammonia fractional abundance, $\chi(NH_3)$. Benson & Myers (1983) studied the relative abundance of CO and $NH_3$ in nearby dark cloud and inferred $\chi(NH_3)$ between $2\times 10^{-7}$ and $3\times 10^{-8}$. Harju et al. (1993) conducted ammonia observations towards 43 massive star forming regions in Orion and Cepheus and found that in nearby regions, i.e., Orion and Cepheus near, $\chi(NH_3)$ varies between $10^{-8}$ and $5 \times 10^{-8}$, while for distant Cepheus sources (d $\sim$ 3500 pc), $\chi(NH_3)$ is below $10^{-8}$. The latter however was explained to be not inconsistent with the former considering that $\chi(NH_3)$ is determined from the rate $N(NH_3)/N(CO)$ and the $NH_3$ beam-filling factor for distant sources is smaller than that of nearby sources resulting in underestimated values of $\chi(NH_3)$ for the distant Cepheus sources. Moreover, the time variations of $\chi(NH_3)$ have been found to be small, varying between $2 \times 10^{-8}$ and $5 \times 10^{-8}$, during the early stage of protostellar evolution (Bergin & Langer 1997; Langer et al. 2000). Here we adopt a constant $\chi(NH_3)$ of $3 \times 10^{-8}$ to estimate hydrogen column densities and core masses. If we assume that the $\chi(NH_3)$ is relatively stable within an order of magnitude, we estimate that the hydrogen column densities and core masses are uncertain to a factor of 4. While the hydrogen column densities and mass estimates based on the constant $\chi(NH_3)$ may be considerably in error in individual cases, we believe that they are reasonably accurate when applied to an ensemble of sources. Consequently, we believe that the larger $N(NH_3)$ of the high-luminosity maser regions compared to those of the low-luminosity maser regions reflects their larger molecular hydrogen column densities. Further, the combination of larger column densities and lower densities for the high-luminosity masers suggests that they are associated with larger cores.

### 4.4. Core Size and Mass

It is known that as protostars evolve, the surrounding dense envelopes, which are the reservoirs for the central objects, evolve simultaneously. Thus differences in the physical properties of the envelopes such as size and mass may shed some light on differences in the evolution of the central young stellar objects (YSOs).

For cores associated with the low-luminosity masers, core sizes range from 0.1 to 0.6 pc, with a mean value $\sim 0.3$ pc. For the high-luminosity masers, only 4 sources were mapped with core sizes ranging from 0.4 to 0.7 pc, the mean value being $\sim 0.6$ pc. Considering the small sample size and our poor angular resolution, this does not provide a statistically significant argument for high-luminosity maser cores having larger sizes than their low-luminosity counterparts. However, as mentioned previously, if we believe the conclusion that high-luminosity maser regions have larger column densities but smaller densities, then one would expect them to be associated with larger cores.

The core masses associated with the low-luminosity masers range from 5 to 142 $M_\odot$, with a mean value of 50 $M_\odot$. For the high-luminosity masers, core masses range from 190 to 650 $M_\odot$, with a mean value of 470 $M_\odot$, which are approximately an order of magnitude larger than their low-luminosity counterparts. It should be noted that the mass of the core associated with the low-luminosity maser 206.54-16.36 is only 5 $M_\odot$, which is too low to form a massive star ($> 8\ M_\odot$). For this source, Mezger et al. (1992) estimated a mass of 10 $M_\odot$ using continuum emission between 350 and 1300 $\mu$m. The uncertainty in our core masses are primarily from the uncertainty in the ammonia abundance. As discussed in §4.3, $\chi(NH_3)$ in Orion was found to vary between $10^{-8}$ and $5 \times 10^{-8}$ within the Orion molecular cloud (Harju et al. 1993). We have used an constant abundance value of $3 \times 10^{-8}$ to calculate the core masses. However, even considering the uncertainties in the mass determination, it is unlikely that 206.54-16.36 will end up being a massive star. Consequently, some weak methanol masers are likely associated with YSOs of intermediate mass. This is consistent with the lower protostellar mass limit of $\sim 3\ M_\odot$ derived by Minier et al. (2003) based on their search for 6.7 GHz methanol masers towards low-mass YSOs.

## 4.5. Outflow Properties

Outflows are significant signatures of star formation and fairly common towards high-mass young stellar objects (Zhang et al. 2001; Beuther et al. 2002b; Xu et al. 2006b). Physical properties of outflows, such as mass, mechanical force and luminosity, are statistically correlated with the properties of the central YSO such as core mass and bolometric luminosity (Shepherd & Churchwell 1996; Beuther et al. 2002b; Wu et al. 2004).

Comparing the properties of the line wings of the CO (1-0) spectra, the high-luminosity maser regions have a larger full width (measured at the 0.1 K level), ranging from 24 to 43 km s$^{-1}$ with a mean value of 34 km s$^{-1}$, than those from the low-luminosity maser regions, where they range from 17 to 25 km s$^{-1}$, with a mean value of 22 km s$^{-1}$. In addition, the CO (1-0) maps show the sizes of outflows from high-luminosity maser regions to be larger, ranging from 0.68 to 0.89 pc with a mean value of 0.84 pc, than those from low-luminosity maser regions, where they range from 0.22 to 0.68 pc, with a mean value of 0.49 pc. There is no obvious distinction of mass in the outflow between the high-luminosity and low-luminosity maser regions. However, as the sample size in both groups is limited, the conclusions obtained here need to be corroborated using a larger sample.

## 4.6. Distance Effect

The primary criterion used to select the maser samples is the isotropic maser luminosity. Ideally, in a comparative analysis, the distances to the sources in both samples should be similar. Unfortunately, it is difficult to find adequate numbers of either low- or high-luminosity 6.7-GHz masers at similar distances. On average, the high-luminosity masers are relatively farther compared to the low-luminosity sources. Thus careful attention should be paid to whether any observational biases are introduced on this account before we obtain the final conclusions.

Firstly, we examine the diagrams of NH$_3$ (1,1) excitation temperatures and kinetic temperatures versus distance. Although the kinetic temperatures and excitation temperatures of the NH$_3$ (1,1) transition from the high-luminosity masers are relatively larger and smaller respectively than those of the low-luminosity maser regions, they have no distance dependence.

In addition, there is also no systematic trend in densities and column densities as a function of distance. The densities of molecular hydrogen were calculated from the kinetic and excitation temperatures of the NH$_3$ (1,1) line, which have no dependence on distance. Hence, it is natural that the molecular densities do not show any dependence on distance.

On the other hand, column densities can decrease due to beam dilution at larger distances. However, our results indicate that the high-luminosity masers on average have larger column densities than their low-luminosity counterparts showing that we are not affected by beam dilution.

### 4.7. Origin of these distinctions

It is interesting that we find systemic differences between high- and low-luminosity 6.7-GHz methanol masers. Theoretically, the luminosities of the masers are determined by the physical environments of the exciting YSOs. (Cragg et al. 2005) found that too large density can quench the masers while regions with larger specific column density can give birth to luminous masers. Our findings that regions associated with high-luminosity 6.7-GHz masers have larger column density but smaller density give a good support to their theory. We find that cores associated with high-luminosity 6.7-GHz masers are larger and more massive that the cores associated with low-luminosity masers. A natural interpretation is that masers with higher luminosity are associated with stars with larger masses, as massive cores tend to give birth to massive stars. One case supports this point of view is the source 206.54-16.36. This maser with lowest luminosity in our sample is associated with NGC2024 FIR 4. The ammonia core surrounding this masers is only 5 $M_\odot$ (Table 6). Moore et al. (1995) estimated the bolometric luminosity of the central star to be 25 $L_\odot$, corresponding to a low mass main sequence star.

Another interpretation is that the 6.7-GHz masers with higher luminosity are generated at a more evolved stage of massive star formation compared with their low-luminosity counterparts. The evolved stars with increasing masses tend to have larger gravitational spheres and accrete more lower density cloud gases from their individual accretion domains around them. Moreover, the more evolved stars release much more energy than less evolved stars which can also explain the larger linewidth for the high-luminosity maser regions. Breen et al. (2010) found that more luminous 6.7-GHz methanol masers are more likely to associated with OH masers which supports this evolution scenario.

In principle, if we have accurate evolutionary tracks of massive protostars and also obtain the luminosity and effective temperature of the stars pumping the masers, we can disentangle the two factors. However, in practice, this is difficult considering our relatively poor understanding of evolution of massive YSOs, which continue to accrete mass after reaching the zero age main sequence. It is also likely that both the stellar mass and evolution have effects on the maser luminosity.

## 5. Conclusions

To investigate distinctions between low- and high-luminosity 6.7-GHz masers, we performed multi-line mapping observations, including $H_3$ (1,1), (2,2), (3,3), (4,4) and $^{12}CO$ (1-0) towards a sample of 9 low-luminosity 6.7-GHz masers and $^{12}CO$ (1-0) observations towards a sample of 8 high-luminosity 6.7-GHz masers with pre-existing $NH_3$ data. We observe that

1. High-luminosity maser regions have lower $NH_3$ (1,1) excitation temperatures, $\sim < 4K$, than that of low-luminosity maser regions.

2. High-luminosity maser regions have $\sim 3$ times larger column densities but smaller densities than that of low-luminosity maser regions.

3. Molecular cores associated with high-luminosity masers are $\sim 10$ times more massive than those associated with low-luminosity masers. In addition the cores bearing high-luminosity masers tend to be larger.

4. Outflows in high-luminosity masers regions have line wings wider by $\sim 10$ km s$^{-1}$ and larger sizes than the outflows in low-luminosity maser regions.

5. Strong correlations are found between 6.7-GHz maser luminosity and line width of $NH_3$ (1,1) and (2,2), indicating that internal motions of high-luminosity maser regions are more dynamic than that of low-luminosity maser regions.

We however caution that the statistical uncertainty is large considering the small number of sources in our study. Consequently, this work has to be repeated towards a much larger sample to confirm the universality of the conclusions above.

This work was supported by the Chinese NSF through grants NSF 10673024, NSF 10733030, NSF 10703010 and NSF 10621303, and NBRPC (973 Program) under grant 2007CB815403.

Table 1: List of low-luminosity 6.7-GHz masers. Column 1 are the names of the methanol masers, Cols. 2, 3 are equatorial coordinates. Peak flux density, distance and maser luminosity are listed in Cols. 4−6, respectively. Other name and reference are listed in Cols. 7 and 8, respectively.

| source name | RA(J2000) ($^h$ $^m$ $^s$) | DEC(J2000) ($^\circ$ $'$ $''$) | $S_{peak}$ (Jy) | D (kpc) | Luminosity ($L_\odot$) | other name | Ref. |
|---|---|---|---|---|---|---|---|
| 106.80+5.31 | 22 19 18.3 | +63 18 48 | 0.5 | 0.9[5] | 7.0E-10 | S 140 | 14 |
| 111.25-0.77 | 23 16 09.7 | +59 55 29 | 4.0 | 3.5[9] | 8.5E-08 | IRAS 23139+5939 | 13 |
| 121.24-0.34 | 00 36 47.358 | +63 29 02.18 | 10 | 0.85[16] | 1.3E-08 | L 1287 | 10,15 |
| 133.72+1.22 | 02 25 41.9 | +62 06 05 | 5 | 2.3[6] | 4.6E-08 | W3 IRS5 | 12 |
| 183.35-0.59 | 05 51 10.8 | +25 46 14 | 19 | 2.1[7] | 1.5E-07 | IRAS 05480+2545 | 12 |
| 188.80+1.03 | 06 09 07.8 | +21 50 39 | 4.8 | 2[1] | 3.3E-08 | AFGL 5182 | 13 |
| 189.03+0.78 | 06 08 40.671 | +21 31 06.89 | 17 | 1.5[8] | 6.6E-08 | AFGL 6466 | 3,15 |
| 189.78+0.34 | 06 08 35.28 | +20 39 06.7 | 15 | 1.5[★] | 5.8E-08 | S 252A | 3,4 |
| 206.54-16.36 | 05 41 44.15 | -01 54 44.9 | 1.48 | 0.415[1] | 4.4E-10 | NGC 2024 | 11 |

References for sources and distances:
(1) Anthony-Twarog 1982;(2) Carpenter et al. 1995; (3) Caswell et al. 1995; (4) Caswell 2009;
(5) Crampton & Fisher 1974; (6) Georgelin & Georgelin 1976; (7) Hughes & Macleod 1993;
(8) Humphreys 1978; (9) Larionov et al. 1999; (10) Macleod et al. 1998;
(11) Minier et al. 2003; (12) Slysh et al. 1999; (13) Szymczak & Kus 2000;
(14) Xu et al. 2008; (15) Xu et al. 2009b; (16) Yang et al. 1991.

★Heliocentric kinematic distance.

Table 2: List of high-luminosity 6.7-GHz masers. Columns are same as that of Table 1.

| source name | RA(J2000) ($^h$ $^m$ $^s$) | DEC(J2000) ($^\circ$ $'$ $''$) | $S_{peak}$ (Jy) | D (kpc) | Luminosity ($L_\odot$) | other name | Ref. |
|---|---|---|---|---|---|---|---|
| 9.62+0.20 | 18 06 14.6568 | -20 31 31.670 | 5090 | $5.2^8$ | 2.38E-04 | IRAS 18032-2032 | 2,8 |
| 12.68-0.18 | 18 13 54.75 | -18 01 46.6 | 544 | $6.6^4$ | 4.10E-05 | W33 B | 2,3,5 |
| 35.20-1.74 | 18 58 13.0517 | +01 40 35.674 | 560 | $3.27^{14}$ | 1.04E-05 | IRAS 18592+0108 | 2,5,14 |
| 49.49-0.39 | 19 23 43.949 | +14 30 34.44 | 850 | $5.1^{12}$ | 3.83E-05 | W51 e | 2,12 |
| 111.54+0.78 | 23 13 45.3622 | +61 28 10.507 | 296 | $2.65^7$ | 3.60E-06 | NGC 7538 | 5,7 |
| 133.94+1.04 | 02 27 03.8 | +61 52 25 | 3880 | $1.95^{11}$ | 2.55E-05 | W3(OH) | 5 |
| 173.49+2.42 | 05 39 13.059 | +35 45 51.29 | 256 | $2.3^1$ | 2.35E-06 | S 231 | 5,6,10 |
| 188.95+0.89 | 06 08 53.342 | +21 38 29.09 | 495 | $2^9$ | 3.43E-06 | AFGL5180(S 252) | 2,13 |

Table 3: NH$_3$ (1,1) and (2,2) spectral parameters of high-luminosity 6.7-GHz masers. Column 1 is the source name which is named after Galactic coordinates, Cols. 2-4 are main beam brightness temperature, the LSR velocity and the line width (1,1) line, respectively. Col. 5 is the optical depth of the main hyperfine component. Cols. 6−8 are same as Cols. 2−4 but for (2,2) line. Col. 9 is reference.

| source name | T$_B$(1,1) (K) | V$_{LSR}$ (km s$^{-1}$) | $\Delta v$(1,1) (km s$^{-1}$) | $\tau(1,1,m)$ | T$_B$(2,2) (K) | V$_{LSR}$ (km s$^{-1}$) | $\Delta v$(2,2) (km s$^{-1}$) | Ref. |
|---|---|---|---|---|---|---|---|---|
| 9.62+0.20 | 3.90(28) | 4.4 | — | 1.65 | 2.37(1) | 4.3(1) | 4.5(2) | 1 |
| 12.68-0.18 | 3.0(5) | 55.7 | 3.3 | 2.4 | 2.5(4) | 55.7 | 2.9 | 2 |
| 35.20-1.74 | 0.88(16) | 44.5 | — | 0.8 | 0.53(03) | 44.3 | 4.2(3) | 1 |
| 49.49-0.39 | 2.3(2) | 56.9(1) | 8.8(2) | 0.75(10) | 2.5(2) | 56.9(1) | 8.3(2) | 3 |
| 111.54+0.78 | 1.2 | -56.9(1) | 2.9(1) | 0.65 | 0.9 | -56.8(1) | 3.2(1) | 6 |
| 133.94+1.04 | 1.2(1) | -47.0 | 4.1(1) | 1.7(2) | 1.1 | -46.7 | 4.9(1) | 4 |
| 173.49+2.42 | 2.43(10) | -16.37(02) | 2.16(06) | 0.97(13) | 1.61(22) | -16.24(03) | 1.90(11) | 7 |
| 188.95+0.89 | 1.62(17) | 3.15 | 2.68 | — | 1 | — | 3.15 | 5 |

References for sources and distances:

(1) Churchwell et al. 1990; (2) Ho et al. 1981; (3) Mauersberger et al. 1985;

(4) Mauersberger et al. 1988; (5) Schreyer et al. 1996; (6) Wilson et al. 1983;

(7) Zinchenko et al. 1997.

Table 4: Observation Parameters.

| Translation | $\nu_{rest}$ (GHz) | HPBW ($''$) | Bandwidth (MHz) | $\Delta \nu_{res}$ (km s$^{-1}$) | 1$\sigma$ rsm[a] (K) |
|---|---|---|---|---|---|
| NH$_3$(1, 1) | 23.6944955 | 40 | 80 | 0.5 | 0.15 |
| NH$_3$(2, 2) | 23.7226336 | 40 | 80 | 0.5 | 0.12 |
| NH$_3$(3, 3) | 23.8701296 | 40 | 80 | 0.5 | 0.12 |
| NH$_3$(4, 4) | 24.1394169 | 40 | 80 | 0.5 | 0.12 |
| $^{12}$CO (1-0) | 115.271204 | 58 | 145 | 0.37 | 0.10 |

[a] typical value in the scale of main beam temperature.

Table 5: NH$_3$ observed parameters at peak position of the cores. Columns 2 is the number of the core. Cols. 3–6 are main beam brightness temperature, the LSR velocity, the line width and the integrated intensity of the main hyperfine component of (1,1) line, respectively. Col. 7 is the optical depth of the main hyperfine component. Cols. 8–11 and Cols. 12–15 are same as Cols. 3–6 but for (2,2) and (3,3) lines.

| Region | Core | $T_B(1,1)$ (K) | $V_{LSR}$ (km s$^{-1}$) | $\Delta v(1,1)$ (km s$^{-1}$) | $\int T_B(1,1)dv$ (K km s$^{-1}$) | $\tau(1,1,m)$ | $T_B(2,2)$ (K) | $V_{LSR}$ (km s$^{-1}$) | $\Delta v(2,2)$ (km s$^{-1}$) | $\int T_B(2,2)dv$ (K km s$^{-1}$) | $T_B(3,3)$ (K) | $V_{LSR}$ (km s$^{-1}$) | $\Delta v(3,3)$ (km s$^{-1}$) | $\int T_B(3,3)dv$ (K km s$^{-1}$) |
|---|---|---|---|---|---|---|---|---|---|---|---|---|---|---|
| 106.80+5.31 | 1 | 5.29 | -7.11(0.01) | 1.32(0.02) | 9.42 | 1.06(0.09) | 3.28 | -7.13(0.01) | 1.40(0.02) | 3.29 | 1.16 | -7.19(0.04) | 1.80(0.94) | 1.22 |
| 106.80+5.31 | 2 | 2.36 | -6.90(0.03) | 2.43(0.05) | 4.29 | 1.13(0.05) | 1.61 | -6.96(0.03) | 2.49(0.08) | 4.28 | 1.27 | -6.87(0.07) | 2.51(0.20) | 3.40 |
| 111.25-0.77 | 1 | 0.87 | -44.04(0.11) | 1.88(0.31) | 0.87 | 0.10(1.00) | 0.45 | -44.06(0.07) | 1.75(0.19) | 0.45 | 0.24 | -44.18(0.23) | 4.24(0.62) | 0.24 |
| 121.24-0.34 | 1 | 4.74 | -17.29(0.02) | 1.62(0.04) | 6.20 | 0.71(0.14) | 1.95 | -17.29(0.02) | 1.69(0.07) | 1.98 | 0.78 | -17.33(0.08) | 3.00(0.22) | 0.74 |
| 183.35-0.59 | 1 | 3.07 | -9.61(0.02) | 1.45(0.05) | 5.80 | 1.44(0.20) | 1.26 | -9.58(0.03) | 1.75(0.07) | 1.38 | 0.28 | -9.64(0.11) | 1.44(0.25) | 0.33 |
| 188.80+1.03 | 1 | 0.99 | -0.54(0.03) | 2.67(0.08) | 2.83 | 0.10(0.06) | 0.60 | -0.34(0.06) | 2.83(0.14) | 2.72 | 0.42 | -0.16(0.08) | 2.58(0.18) | 1.16 |
| 188.80+1.03 | 2 | 1.07 | -1.39(0.03) | 2.07(0.08) | 2.83 | 0.80(0.18) | 0.28 | -1.30(0.12) | 1.74(0.45) | 0.31 | — | — | — | — |
| 189.03+0.78 | 1 | 1.46 | 2.44(0.05) | 1.86(0.15) | 1.75 | 0.35(0.36) | 0.80 | 2.51(0.05) | 1.44(0.01) | 1.12 | 0.25 | 2.35(0.24) | 3.84(0.56) | 0.26 |
| 189.03+0.78 | 2 | 1.77 | 1.35(0.02) | 1.55(0.06) | 2.55 | 0.64(0.16) | 0.86 | 1.43(0.05) | 1.61(0.13) | 0.93 | — | — | — | — |
| 189.78+0.34 | 1 | 2.50 | 8.95(0.02) | 1.89(0.07) | 3.36 | 0.58(0.16) | 1.04 | 8.97(0.06) | 2.09(0.14) | 1.12 | 0.57 | 9.10(0.11) | 2.81(0.26) | 0.53 |
| 189.78+0.34 | 2 | 3.44 | 7.38(0.02) | 1.87(0.05) | 4.31 | 0.56(0.10) | 0.95 | 7.45(0.05) | 1.40(0.13) | 1.91 | 0.43 | 7.65(0.15) | 3.41(0.46) | 0.34 |
| 189.78+0.34 | 3 | 1.96 | 7.61(0.04) | 2.24(0.12) | 2.44 | 0.53(0.25) | 0.57 | 7.68(0.12) | 1.32(0.20) | 1.45 | — | — | — | — |
| 206.54-16.36 | 1 | 2.38 | 10.12(0.03) | 1.22(0.07) | 3.58 | 0.63(0.32) | 1.76 | 10.10(0.06) | 1.24(0.16) | 3.61 | 1.55 | 10.06(0.07) | 2.06(0.23) | 1.67 |
| 206.54-16.36 | 2 | 3.08 | 10.98(0.02) | 2.02(0.05) | 4.65 | 0.91(0.12) | 2.29 | 10.92(0.03) | 2.17(0.10) | 3.00 | 1.39 | 10.88(0.05) | 2.89(0.12) | 1.52 |

Table 6: Physical quantities of cores associated with low-luminosity 6.7-GHz methanol masers. Cols. 3, 4 are equatorial coordinates, Cols. 5 are angular extensions of the major and minor axes of the core assuming a spherical geometry. Cols. 6 is the nominal core size. Rotation temperature of the (2, 2) and (1, 1) transitions, kinetic temperature and excitation temperature of the (1, 1) line are listed in Cols. 7−9, respectively. Molecular hydrogen density, $NH_3$ column density and mass are listed in Cols. 10−12, respectively. Mean and median values of the quantities for the low-luminosity maser regions are listed in the last two rows.

| Region Name | Core | R.A. ($^h$ $^m$ $^s$) | DEC. ($^\circ$ $^\prime$ $^{\prime\prime}$) | angular ($^{\prime\prime}$,$^{\prime\prime}$) | size (pc) | $T_{R21}$ (K) | $T_{Kin}$ (K) | $T_{ex}$(1,1) (K) | $n_{H_2}$ ($\times 10^4 cm^{-3}$) | N($NH_3$) ($\times 10^{14} cm^{-2}$) | M ($M_\odot$) |
|---|---|---|---|---|---|---|---|---|---|---|---|
| 106.80+5.31 | 1 | 22:19:23.0 | +63:19:19 | (80,60) | 0.20 | 21.2(0.8) | 29.8(1.6) | 10.2(0.9) | 2.0(0.2) | 7.3 | 25.3 |
| 106.80+5.31 | 2 | 22:19:16.0 | +63:18:45 | (60,60) | 0.15 | 21.2(0.8) | 29.8(1.6) | 10.2(0.9) | 2.0(0.2) | 4.5 | 8.8 |
| 111.25-0.77 | 1 | 23:16:10.0 | +59:55:30 | (60,60) | 0.59 | 21.8(2.7) | 31.0(5.0) | 11.1(1.0) | 2.3(2.0) | 1.3 | 39.2 |
| 121.24-0.34 | 1 | 00:36:47.8 | +63:28:57 | (100,65) | 0.26 | 17.5(0.6) | 23.0(1.0) | 11.4(0.4) | 3.1(0.1) | 9.0 | 52.7 |
| 183.35-0.59 | 1 | 05:51:10.5 | +25:46:05 | (80,50) | 0.40 | 16.0(0.4) | 19.5(0.8) | 6.8(0.2) | 1.1(0.03) | 10.3 | 142.8 |
| 188.80+1.03 | 1 | 06:09:06.5 | +21:50:54 | (80,75) | 0.54 | 23.3(2.0) | 34.2(4.1) | 13.1(5.0) | 2.8(2.0) | 5.4 | 55.6 |
| 188.80+1.03 | 2 | 06:09:12.0 | +21:51:11 | (60,50) | 0.47 | 14.5(0.2) | 17.4(0.2) | 4.7(0.1) | 0.5(0.1) | 6.0 | 114.8 |
| 189.03+0.78 | 1 | 06:08:39.9 | +21:31:00 | (80,45) | 0.26 | 20.9(1.0) | 29.2(1.8) | 7.7(2.3) | 1.2(0.4) | 6.0 | 35.1 |
| 189.03+0.78 | 2 | 06:08:46.5 | +21:31:58 | (70,70) | 0.21 | 18.9(0.3) | 25.5(0.6) | 6.6(0.4) | 0.9(0.1) | 4.3 | 16.5 |
| 189.78+0.34 | 1 | 06:08:35.5 | +20:39:00 | (60,45) | 0.29 | 17.9(0.2) | 23.6(0.3) | 8.4(0.6) | 1.6(0.2) | 6.3 | 45.9 |
| 189.78+0.34 | 2 | 06:08:40.5 | +20:37:57 | (100,80) | 0.50 | 15.2(0.1) | 18.3(0.1) | 10.3(0.3) | 3.1(0.2) | 8.0 | 173.3 |
| 189.78+0.34 | 3 | 06:08:42.5 | +20:36:30 | (70,50) | 0.25 | 15.6(0.2) | 18.9(0.3) | 7.2(1.0) | 1.3(0.4) | 6.3 | 34.11 |
| 206.54-16.36 | 1 | 05:41:43.7 | -01:54:20 | (70,60) | 0.12 | 24.8(0.7) | 37.4(1.6) | 8.3(1.3) | 1.3(0.3) | 4.1 | 5.1 |
| 206.54-16.36 | 2 | 05:41:45.3 | -01:56:13 | (70,60) | 0.09 | 24.5(0.3) | 36.7(0.6) | 7.7(0.3) | 1.1(0.1) | 9.0 | 6.3 |
| mean[a] | — | — | — | — | 0.32 | 20.4 | 28.5 | 9.6 | 1.9 | 6.2 | 51.5 |
| median[b] | — | — | — | — | 0.26 | 21.1 | 29.5 | 9.3 | 1.8 | 6.2 | 42.6 |

[a,b]Mean and median values are only for the low-luminosity regions, i.e, core 1, excluding core 2 and 3.

Table 7: Physical quantities of cores associated with high-luminosity 6.7-GHz masers. Cols. 2 is the nominal core size. Rotation temperature of the (2, 2) and (1, 1) transitions, kinetic temperature and excitation temperature of the (1, 1) line are listed in Cols. 3 - 5, respectively. Molecular hydrogen density, $NH_3$ column density and mass are listed in Cols. 6−8, respectively. Mean and median values of these quantities are listed in the last two rows.

| source name | size (pc) | $T_{ex}(1,1)$ (K) | $T_{R21}$ (K) | $T_{kin}$ (K) | $n_{H_2}$ ($\times 10^3 cm^{-3}$) | $N(NH_3)$ ($\times 10^{14} cm^{-2}$) | M ($M_\odot$) |
|---|---|---|---|---|---|---|---|
| 9.62+0.20 | — | 7.5(0.4) | 19.2(0.4) | 26.0(0.8) | 12.2(1.0) | 36.94 | — |
| 12.68-0.18 | 0.53 | 6.0(0.5) | 23.2(0.5) | 33.8(1.0) | 7.2(1.2) | 26.57 | 647.3 |
| 35.20-1.74 | — | 4.3(0.3) | 21.3(2.1) | 29.9(4.2) | 3.4(0.6) | 9.28 | — |
| 49.49-0.39 | — | 7.1(0.1) | 36.5(0.1) | 72.2(0.4) | 8.6(0.1) | 32.30 | — |
| 111.54+0.78 | — | 5.2(0.2) | 25.2(0.5) | 38.3(1.2) | 5.2(0.4) | 6.25 | — |
| 133.94+1.04 | 0.64 | 4.2(0.1) | 28.2(0.2) | 45.7(0.3) | 2.9(0.1) | 18.70 | 633.6 |
| 173.49+2.42 | 0.70 | 6.6(0.2) | 22.2(1.1) | 31.8(2.4) | 8.8(0.2) | 8.85 | 375.7 |
| 188.95+0.89 | 0.44 | 4.9(0.2) | 20.2(0.6) | 27.8(1.2) | 4.8(0.5) | 11.44 | 191.2 |
| mean | 0.6 | 5.7 | 24.5 | 38.2 | 6.6 | 18.79 | 469.5 |
| median | 0.6 | 5.6 | 22.7 | 32.8 | 6.2 | 15.07 | 511.5 |

Table 8: Parameters of outflow line wings. Cols. 2 and 3 are velocity ranges of blue and red wings. Cols. 4 and 5 are integral intensities of emissions of blue and red wings. Col. 6 is full width of line wing at 100 mK level. Cols. The first 6 rows are outflows of low-luminosity maser regions and the last four rows are outflows of strong maser regions.

| source name | $\triangle V_b$ (km s$^{-1}$) | $\triangle V_r$ (km s$^{-1}$) | $S_b$ (K km s$^{-1}$) | $S_r$ (K km s$^{-1}$) | $\triangle$ V(100 mK) (km s$^{-1}$) |
|---|---|---|---|---|---|
| 106.80+5.31 | (-20,-12) | (-4,2) | 24.8 | 23.9 | 23.3 |
| 111.25-0.77 | (-56,-48) | (-39,-36) | 15.1 | 7.1 | 22.2 |
| 121.24-0.34 | (-29,-20) | (-15,-7) | 24.2 | 17.8 | 25.0 |
| 189.03+0.78 | (-8,-1) | (6,12) | 7.0 | 13.1 | 17.5 |
| 189.78+0.34 | (-1,3.5) | (12,21) | 12.0 | 16.6 | 23.5 |
| 206.54-16.36 | — | (14,20) | — | 19.2 | 21.1 |
| 111.54+0.78 | (-75,-65) | (-47,-35) | 12.7 | 22.6 | 43.2 |
| 133.94+1.04 | (-60,-52) | (-41,-40) | 24.2 | 14.4 | 35.1 |
| 173.49+2.42 | (-30,-20) | (-13,0) | 9.2 | 9.2 | 24.5 |
| 188.95+0.89 | (-10,-2) | (8,20) | 6.0 | 12.1 | 31.0 |

Table 9: Outflow properties. Cols 2−4 are the total mass $M_{out}$ in $M_{\odot}$, the momentum $p$ [$M_{\odot}$ km s$^{-1}$] and energy E [$10^{45}$ ergs], respectively. Cols 5 and 6 are outflow size [pc] and characteristic time scale $t$ [$10^4$ yr]. the mass entrainment rate of the molecular $\dot{M}$ [$10^{-5}M_{\odot}$ yr$^{-1}$], the mechanical force $F_m$ [$10^{-4}M_{\odot}$ km s$^{-1}$ yr$^{-1}$] and the mechanical luminosity $L_m$ [$L_{\odot}$] are listed in Cols. 7−9, respectively. The first 6 rows are outflows of low-luminosity maser regions, the last four rows are outflows of strong maser regions.

| source | $M_{out}$ | $p$ | E | Size | $t$ | $\dot{M}_{out}$ | $F_m$ | $L_m$ |
|---|---|---|---|---|---|---|---|---|
| (1) | (2) | (3) | (4) | (5) | (6) | (7) | (8) | (9) |
| 106.80+5.31 | 2.5 | 27.6 | 3.1 | 0.22 | 1.9 | 12.7 | 14.2 | 1.34 |
| 111.25-0.77 | 10.6 | 125.4 | 15.6 | 0.68 | 6.0 | 17.6 | 20.7 | 2.14 |
| 121.24-0.34 | 0.7 | 7.3 | 0.8 | 0.25 | 2.2 | 3.0 | 3.3 | 0.30 |
| 189.03+0.78 | 2.8 | 28.1 | 2.8 | 0.65 | 6.4 | 4.4 | 4.4 | 0.36 |
| 189.78+0.34 | 3.7 | 42.1 | 4.8 | 0.65 | 5.8 | 6.4 | 7.2 | 0.68 |
| 206.54-16.36 | 0.2 | 4.0 | 0.7 | 0.48 | 3.3 | 0.6 | 1.2 | 0.19 |
| 111.54+0.78 | 4.8 | 100.4 | 21.9 | 0.77 | 3.8 | 12.8 | 26.6 | 4.58 |
| 133.94+1.04 | 2.6 | 32.0 | 4.0 | 0.85 | 8.3 | 3.2 | 3.8 | 0.40 |
| 173.49+2.42 | 2.5 | 34.4 | 4.8 | 0.89 | 7.0 | 3.9 | 5.5 | 0.64 |
| 188.95+0.89 | 2.2 | 28.0 | 3.6 | 0.68 | 5.3 | 4.1 | 5.3 | 0.56 |

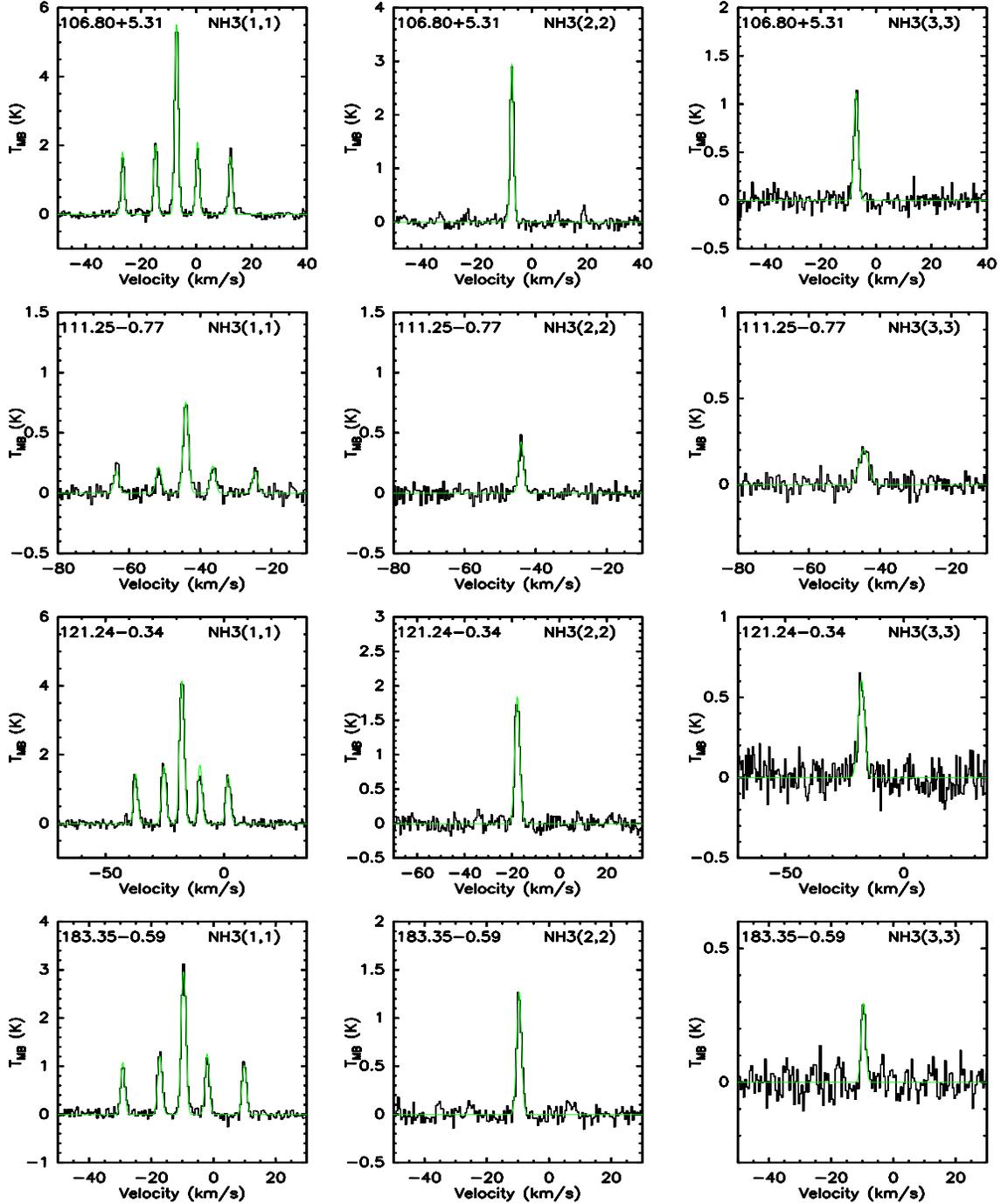

Fig. 1.— NH$_3$ (1,1), (2,2), (3,3) spectra at the peak positions the ammonia cores. The velocities are the radial velocity with respect to the local standard of rest. The y-axis is the main beam temperature. Green lines are fit profiles.

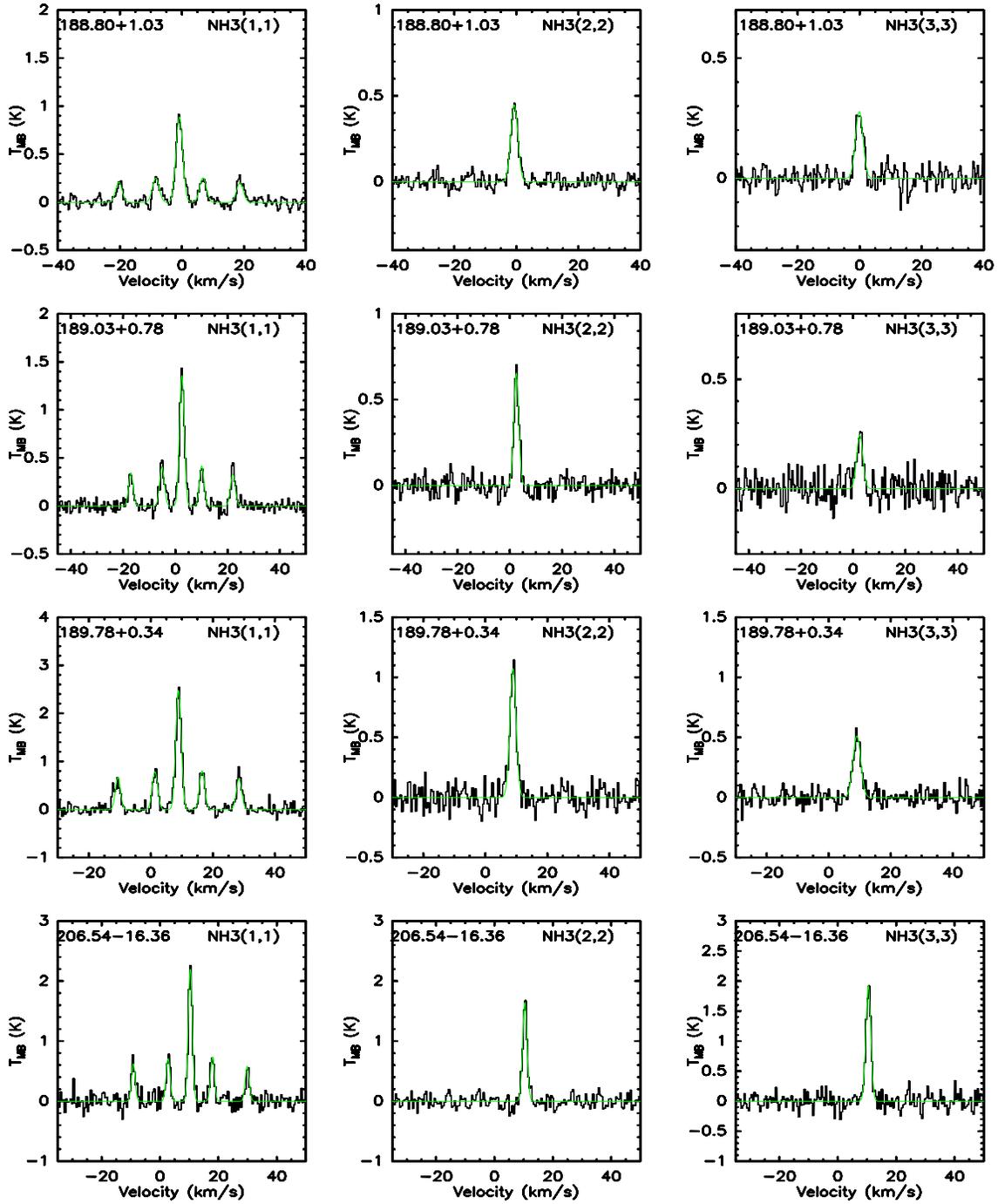

Fig. 1— Continued

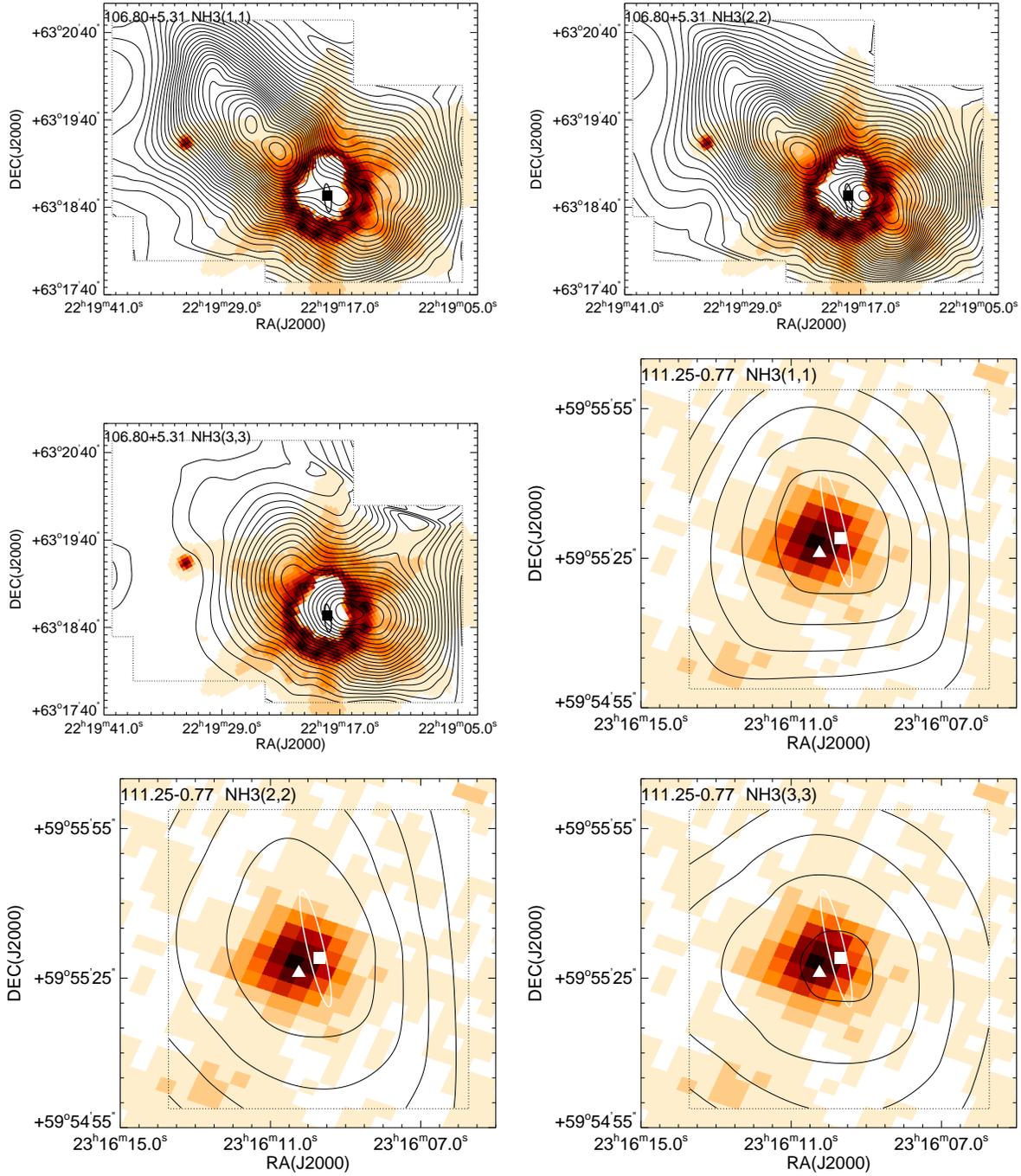

Fig. 2.— NH$_3$ (1,1), (2,2), (3,3) integrated intensity contours of low-luminosity maser regions. The grey scale images are MSX 21 $\mu$m images except for 106.80+5.31, where the grey scale image is MIPS 24 $\mu$m image. The central blank pixels in the 24 $\mu$m image is due to saturation. The open squares, filled triangles and ellipses denote the methanol masers, the MSX point sources and error ellipses of the IRAS point sources in the fields, respectively. Contour levels are from 0.3 K km s$^{-1}$ in steps of 0.3 K km s$^{-1}$ for NH$_3$ (1,1) and 0.15 K km s$^{-1}$ in steps of 0.15 K km s$^{-1}$ for NH$_3$ (2,2) and (3,3).

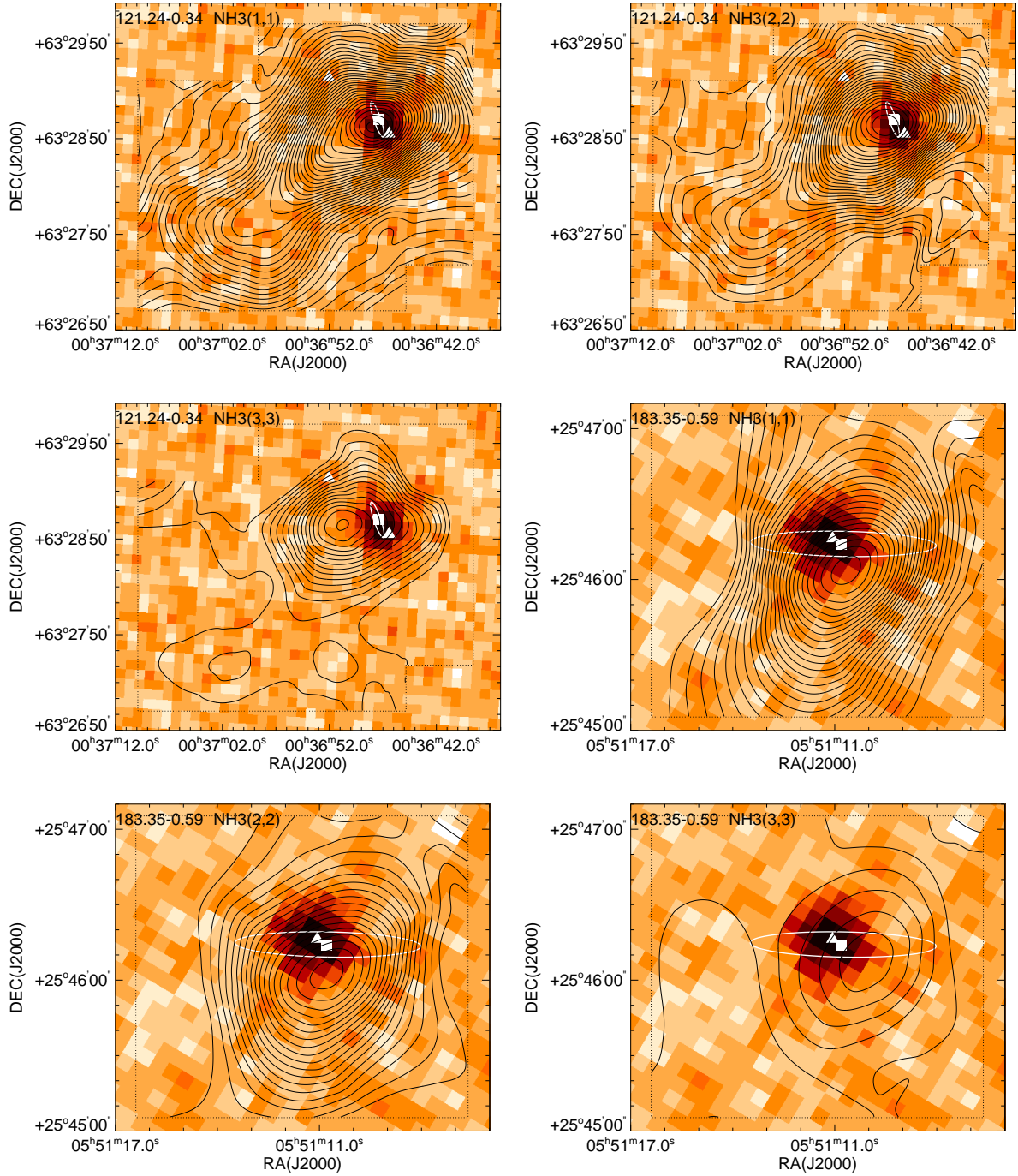

Fig. 2— Continued

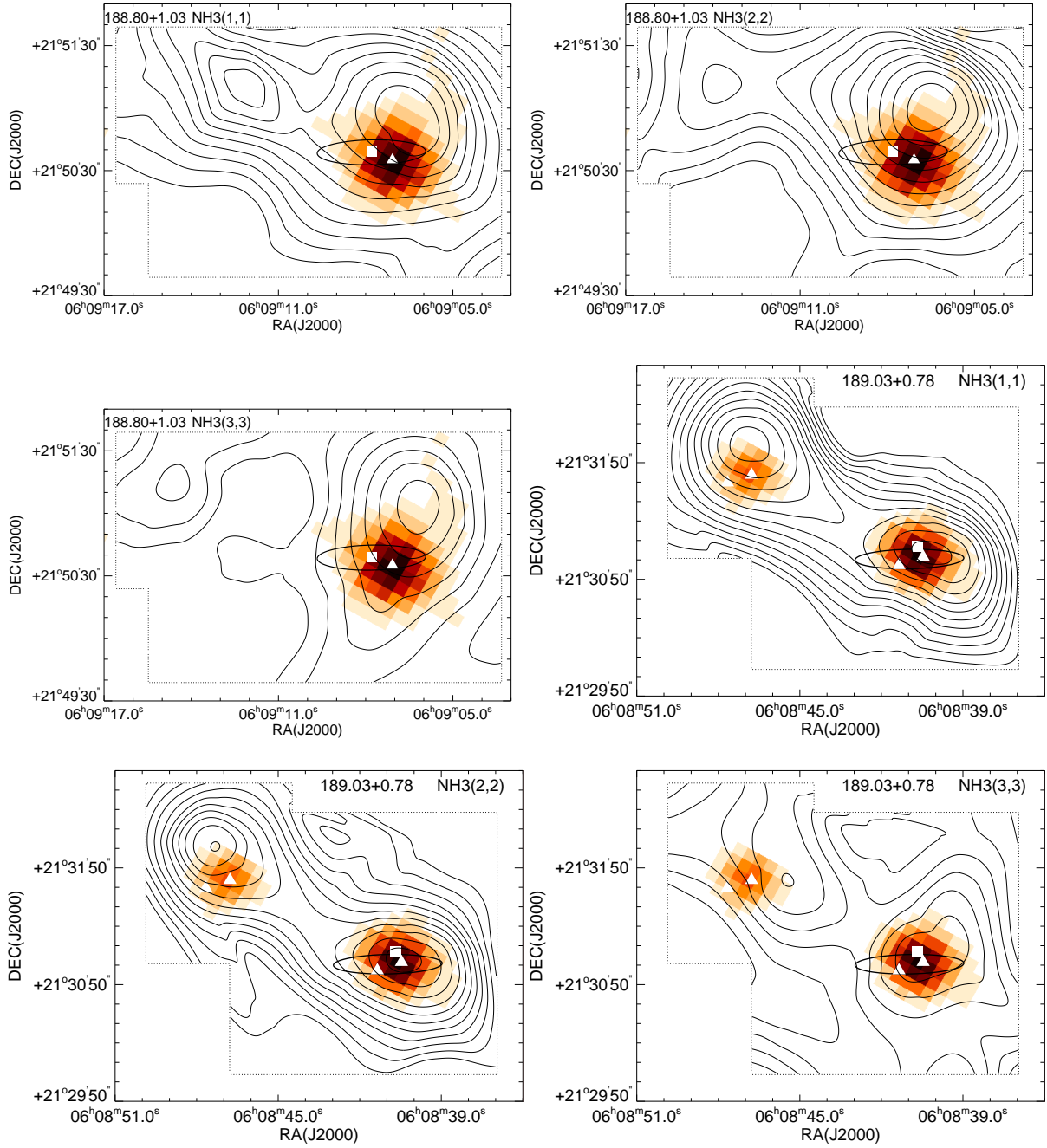

Fig. 2— Continued

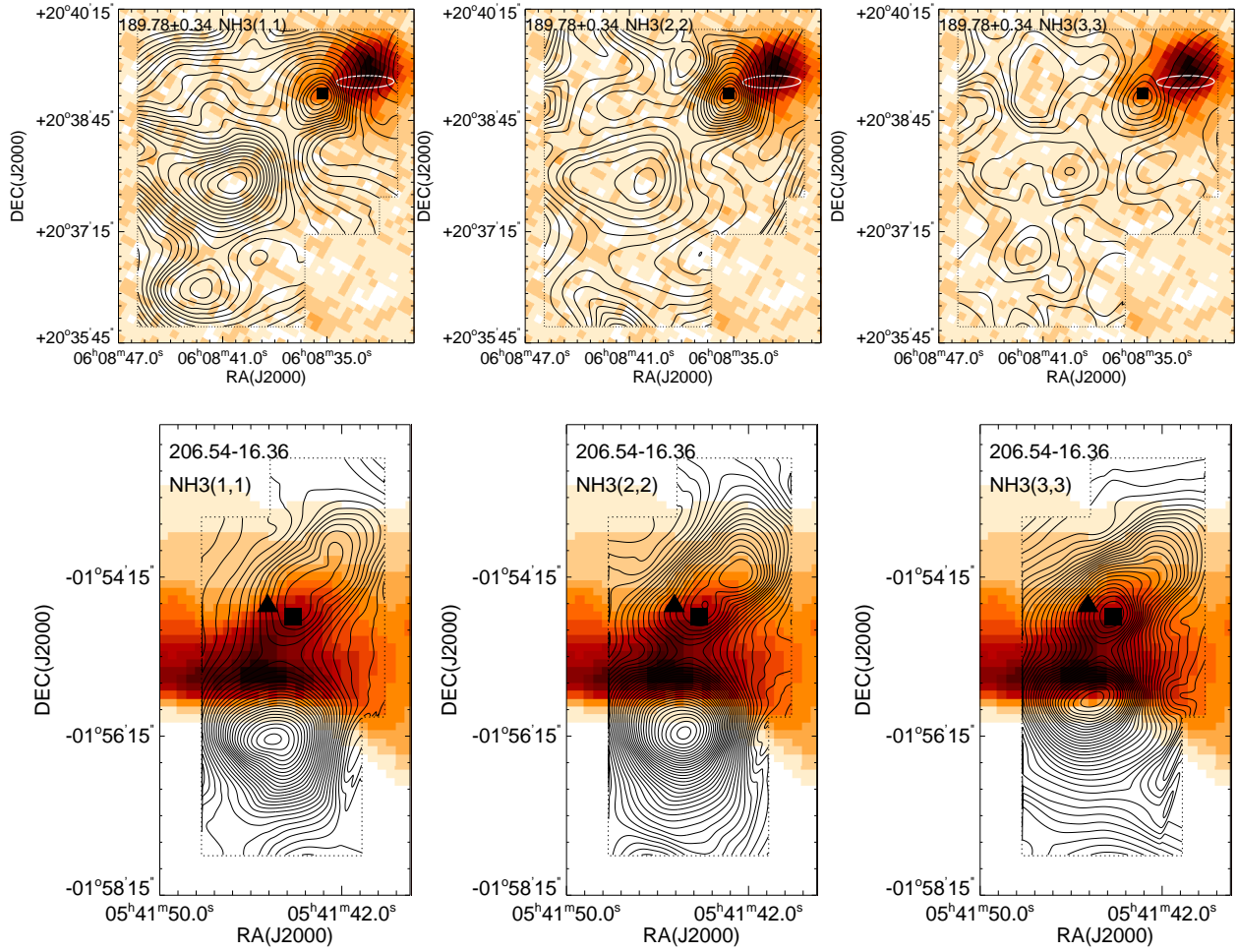

Fig. 2— Continued

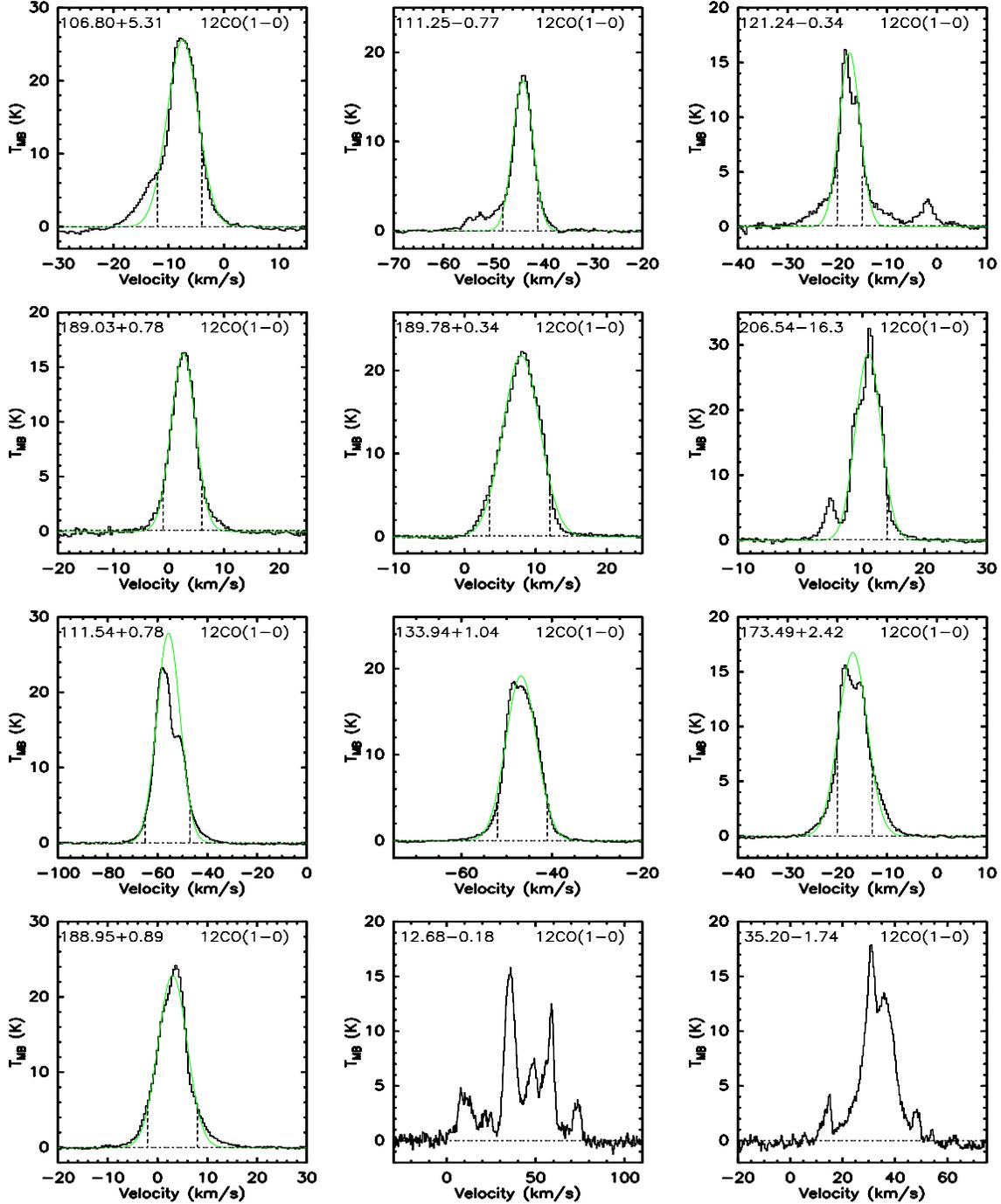

Fig. 3.— $^{12}CO$ (1-0) spectra. The velocities are measured with respect to the LSR, the horizonal dash lines are 100 mK levels, the vertical dash lines are boundaries of blue and red wings, the green profiles are gauss fit profiles.

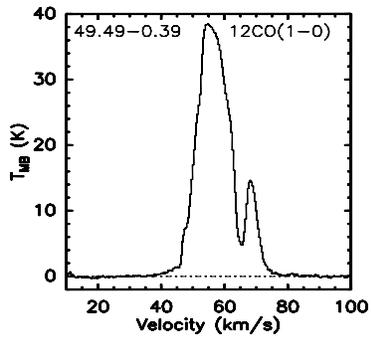

Fig. 3— Continued

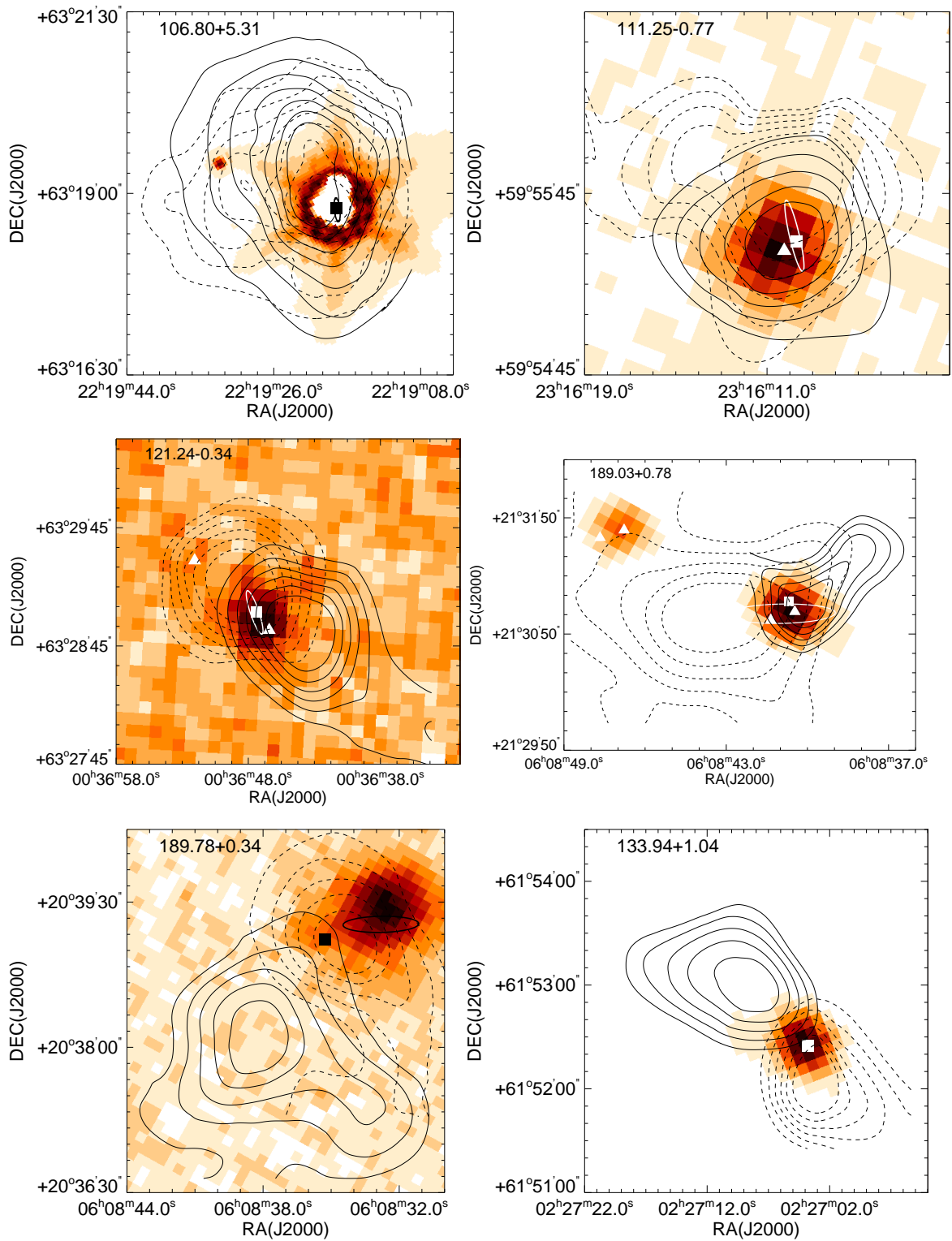

Fig. 4.— Outflows of both high and low-luminosity 6.7-GHz maser regions. The grey scale images and denotations are same as that in Fig. 2. The solid and dashed lines present the blue and red outflow lobes, respectively. The contours are chosen to highlight the most prominent features in each source, usually between 30% and 90% (steps of 10%) of the peak integrated wing intensity.

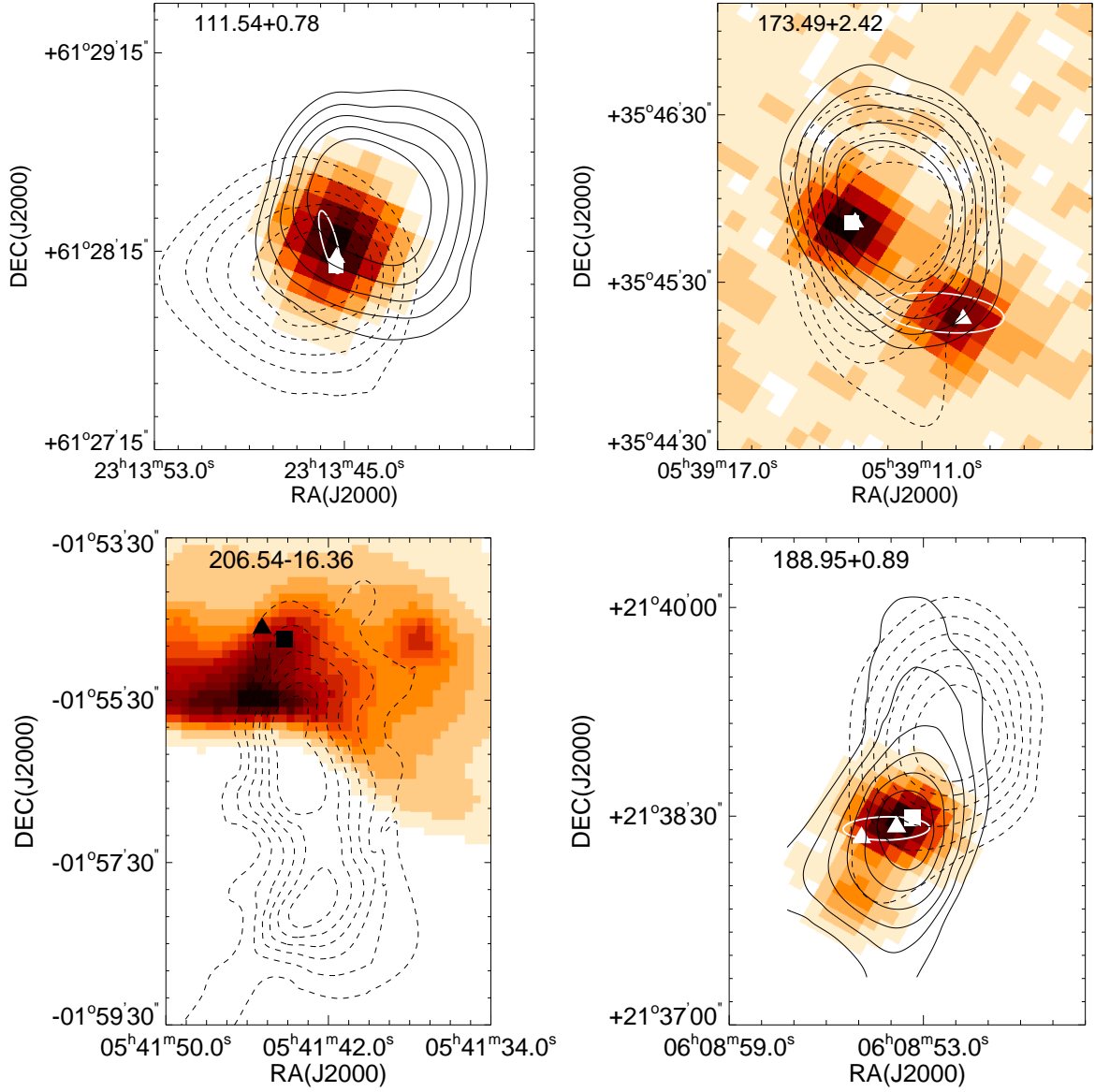

Fig. 4— Continued

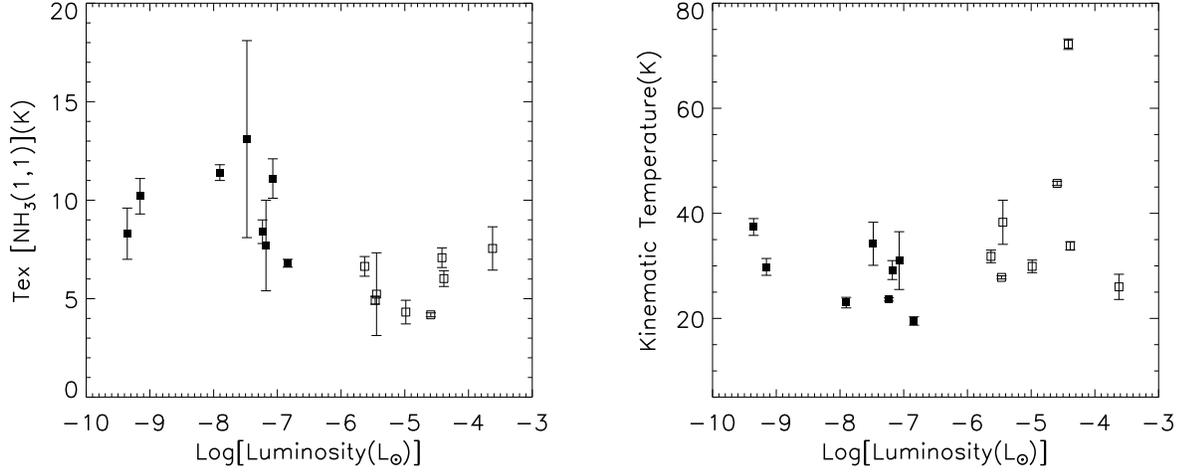

Fig. 5.— Excitation temperature (left panel) and the kinetic temperature (right panel) of the NH₃ (1,1) line as a function of maser luminosity. Filled and open squares indicate low- and high-luminosity masers respectively.

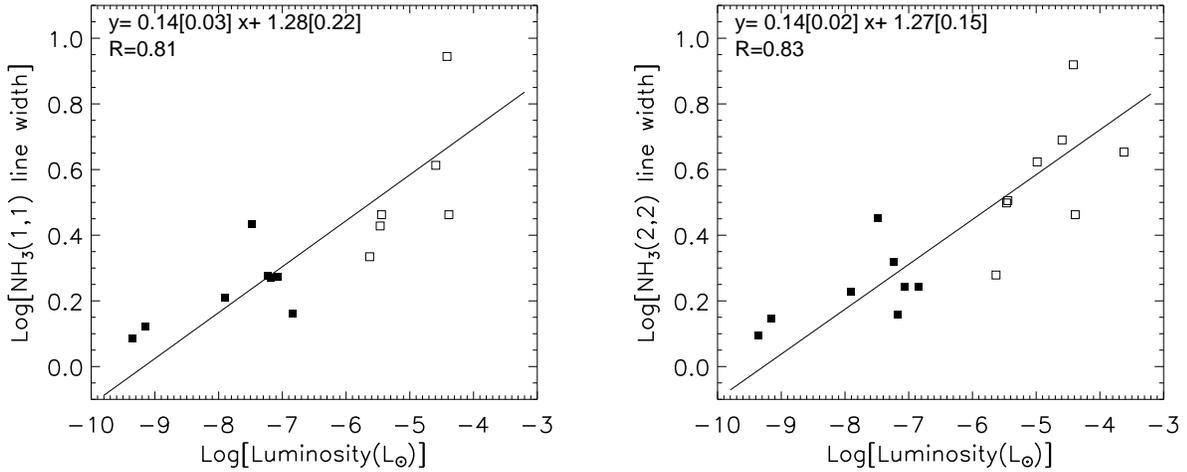

Fig. 6.— Line width of NH₃ (1,1) (left panel) and NH₃ (2,2) (right panel) as a function of maser luminosity. Filled and open squares indicate low- and high-luminosity masers respectively. The solid line is a Least-squares fit of the points, with the fit expression being indicated on the top left corner.

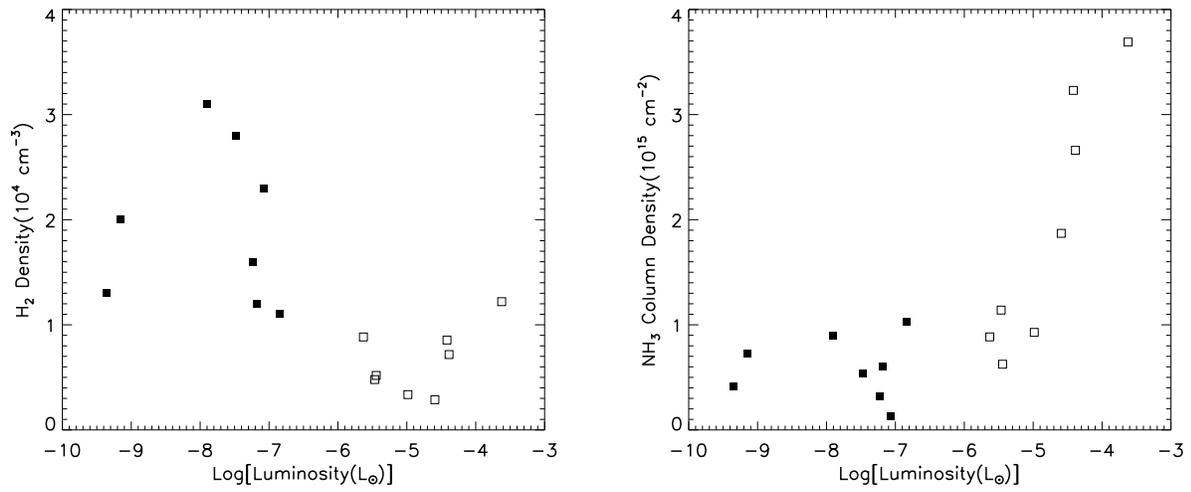

Fig. 7.— $H_2$ volume density (left panel) and $NH_3$ column density (right panel) as a function of maser luminosity. Filled and open squares indicate low- and high-luminosity masers, respectively.